\documentclass[12pt]{article}
\pdfoutput=1
\usepackage{ifpdf}
\usepackage[a4paper,top=3.3cm,width=17cm, bottom=3.4cm, left=2cm]{geometry}
\usepackage{graphicx}
\usepackage{microtype}
\usepackage[utf8x]{inputenc}
\usepackage[T1]{fontenc}
\usepackage{ae}
\usepackage{aecompl}
\usepackage{amssymb}
\usepackage{mathdots}

\usepackage[small,sf,bf,hang]{caption}
\usepackage[english]{babel}
\usepackage{datetime}
\shortdate 

\usepackage{authblk}

\long\def\symbolfootnote[#1]#2{\begingroup%
\def\thefootnote{\fnsymbol{footnote}}\footnote[#1]{#2}\endgroup}

\usepackage[nosort]{cite}

\renewcommand{\vec}[1]{{\bf #1}}

\newcommand{\bea}{\begin{eqnarray}}
\newcommand{\beal}[1]{\begin{eqnarray}\label{#1}}
\newcommand{\eea}{\end{eqnarray}}
\newcommand{\be}{\begin{equation}}
\newcommand{\bel}[1]{\begin{equation}\label{#1}}
\newcommand{\ee}{\end{equation}}
\newcommand{\rf}[1]{Eq.~(\ref{#1})}
\newcommand{\rfs}[1]{Sec.~\ref{#1}}
\newcommand{\rff}[1]{Fig.~\ref{#1}}
\newcommand{\rfr}[1]{Ref.~\cite{#1}}
\newcommand{\f}[2]{\frac{#1}{#2}}

\newcommand{\sym}{${\mathcal N}=4$}

\newcommand{\tpi}{\tau_{\Pi}}

\makeatother

\begin{document}

\title{{\bf Small systems and regulator dependence in relativistic
    hydrodynamics}}
\author[1,2]{Micha\l\ Spali\'nski}
\affil[1]{Physics Department, University of Bia{\l}ystok, Konstantego
  Cio{\l}kowskiego 1L, 15-245 Bia\l ystok, Poland}
\affil[2]{National Center for Nuclear Research, ul. Ho{\.z}a 69, 00-681
  Warsaw, Poland}
\date{}

\maketitle

\thispagestyle{empty}

\begin{abstract}

Consistent theories of hydrodynamics necessarily include nonhydrodynamic modes, which can be viewed as a
regulator necessary to ensure causality. Under many circumstances the choice of regulator is not relevant, but
this is not always the case. In particular, for sufficiently small systems (such as those arising in pA or pp
collisions) such dependence may be inevitable. We address this issue in the context of the modern version of
M\"uller-Israel-Stewart theory of relativistic hydrodynamics. In this case, by demanding that the
nonhydrodynamic modes be subdominant, we find that regulator dependence becomes inevitable only for
multiplicities $dN/dY$ of the order of a few. This conclusion supports earlier studies based on hydrodynamic
simulations of small systems, at the same time providing a simple physical picture of how hydrodynamics can be
reliable even in such seemingly extreme conditions.

\end{abstract}


\section{Introduction}
\label{sec:int}

The modern version of M\"uller-Israel-Stewart theory (MIS) \cite{Muller:1967zza,Israel:1979wp}, which will be
referred to as BRSSS \cite{Baier:2007ix}, is the basic phenomenological tool for understanding the dynamics of
quark-gluon plasma (QGP) produced in heavy ion (AA) collisions. It has recently been found that this theory of
relativistic hydrodynamics works remarkably well also in the case of other processes such as pA or even pp
collisions \cite{CasalderreySolana:2009uk,Bozek:2011if}, which lead to smaller drops of plasma. This raises
the question of why hydrodynamics applies here, and where the limit of its applicability lies
\cite{Shuryak:2013ke,Basar:2013hea,Romatschke:2015gxa,Habich:2015rtj}. The aim of this note is to address this
question in the context of recent advances in our understanding of relativistic hydrodynamics.

The key point is that the factor which signals the emergence of hydrodynamic behaviour in a microscopic theory
such as QCD is the decay of nonhydrodynamic modes. This point has frequently been emphasized in the context of
holographic studies of \sym\ supersymmetric Yang-Mills
plasma~\cite{Kovtun:2005ev,Chesler:2009cy,Heller:2013fn}, but it is valid generally. In particular, it is
valid within for BRSSS theory, which incorporates a particular nonhydrodynamic sector needed for its
self-consistency~\cite{Heller:2015dha}.

Unlike nonrelativistic Navier-Stokes theory, its direct relativistic generalisation \cite{LLfluid} is not
consistent, because it is not causal~\cite{Hiscock:1985zz,Lindblom:1995gp,kostadt}. The only known way to
achieve causality is to include additional modes \cite{Muller:1967zza} beyond the basic hydrodynamic variables
(the energy density and fluid velocity). The simplest example where this works is MIS theory, which adds a
single purely damped nonhydrodynamic degree of freedom. This mode should be thought of as a regulator,
ensuring that the speed of propagation does not exceed the speed of light. Indeed, the speed of propagation of
linear perturbations\footnote{The formula \rf{velo} pertains specifically to the sound channel.} is
\bel{velo}
v = \f{1}{\sqrt{3}} \sqrt{1 + 4 \f{\eta/s}{T \tpi}} \ ,
\ee
where $\eta$ is the shear viscosity, $s$ is the entropy density, $T$ is the effective temperature and $\tpi$
is the relaxation time associated with the nonhydrodynamic mode.\footnote{Only the conformal case will be
discussed explicitly.} This formula (which follows from the sound channel dispersion relation given in
\rf{disprel} below) implies that as long as the relaxation time is sufficiently large
\bel{causal}
T \tpi > 2 \eta/s
\ee
there is no transluminal signal propagation. This is clearly not the case if one tries to eliminate the
relaxation time by taking it to vanish.

The presence of nonhydrodynamic modes is therefore essential for the consistency of the hydrodynamic
description in the relativistic case. The success of relativistic hydrodynamics in describing the dynamics of
QGP created in AA collisions can be ascribed to the exponential decay of these modes, which leads to the fast
emergence of quasiuniversal, attractor behaviour of this system \cite{Heller:2015dha}.

The nonhydrodynamic modes act as a regulator which cannot be removed, but whose effects may or may not be
practically significant in the regime of interest.  It is important to understand when the effects of
nonhydrodynamic modes may be ignored, otherwise one may be studying the physics of the regulator rather than
universal hydrodynamic behaviour. In particular, in the case of a small system it may happen that the
nonhydrodynamic modes do not have time to decay, and hydrodynamic simulations become sensitive to the choice
of the nonhydrodynamic sector -- that is, to the choice of regulator.\footnote{One can think of the regulator
sector as an analogue of the notion of a ``UV-completion'', which arose in the context of effective field
theories.} If this happens, it may be necessary to compare different regulators. Examples of hydrodynamic
theories with a qualitatively different nonhydrodynamic sector were discussed in Ref.~\cite{Heller:2014wfa}.

The statement that hydro works for small systems \cite{Bozek:2011if} means specifically that parameters of
BRSSS theory (or some other variant of MIS theory) can be fitted to describe the data. The point we are making
here is that in some situations this becomes a test of the nonhydrodynamic sector of this theory rather than
of hydrodynamics. This is problematic if one wishes to regard BRSSS as an effective description of QCD plasma.
Implications of this are further discussed in \rfs{sec:sum}.

The question of the domain of validity of hydrodynamics applied to small systems was considered recently in
Refs.~\cite{Romatschke:2015gxa,Habich:2015rtj}, which considered the dependence on the magnitude of second
order terms in the gradient expansion as a measure of systematic error. In cases where this error becomes
significant, the authors concluded that hydrodynamics ceases to be useful. From a theoretical standpoint it is
the decay of the nonhydro modes and not the size of gradient corrections which sets the domain of validity of
hydrodynamics. However, within BRSSS theory the parameter which governs the decay of the nonhydro modes, the
relaxation time, is also responsible for some of the second order terms --- indeed, from a modern perspective
\cite{Baier:2007ix} the MIS relaxation time is just one of a number of second order transport coefficients.
This is clearly the appropriate view when discussing the gradient expansion generated from the hydro equations
of motion. However when solving the equations numerically, the relaxation time should be regarded as a
regularization parameter. If the results depend significantly on the value taken for this parameter, one may
infer that one is not really testing the hydrodynamic sector, but rather the physics of the regulator itself.
Here we follow this logic directly: by comparing the decay rates of the hydro and nonhydro sectors. This leads
to a straightforward analytic argument which results in a simple inequality, whose violation indicates that
nonhydrodynamic modes are not subdominant. This inequality can be phrased either in terms of the size and
temperature of the system, or in terms of the final multiplicity measured. When expressed in terms of local
effective temperature and size the inequality is in fact more general than the context of small systems; it is
a bound on the size of features (such as spikes in the energy density), whose violation implies regulator
dependence.

\section{Dispersion relations in BRSSS theory}
\label{sec:mis}

The BRSSS theory of relativistic hydrodynamics~\cite{Baier:2007ix} is a generalization of the original MIS
theory~\cite{Muller:1967zza,Israel:1979wp}, which includes the full set of transport coefficients allowed by
Lorentz and conformal symmetry (the latter assumption was relaxed in \rfr{Romatschke:2009kr}). The spectrum of
linearized perturbations around equilibrium reveals two types of behavior: hydrodynamic modes whose frequency
vanishes with the wave vector, as well as nonhydrodynamic modes whose frequency approaches a nonzero value at
$k\equiv |\vec{k}|=0$. The imaginary parts of these frequencies determine the decay rates. Formally, at
$k\approx 0$ the hydro modes are long lived, while the nonhydro modes decay exponentially.

The dispersion relations for BRSSS theory, assuming solutions of the form
\be
\delta T \sim \exp\left(-i(\omega t - \vec{k}\cdot\vec{x})\right),\quad
\delta u^\mu \sim \exp\left(-i(\omega t - \vec{k}\cdot\vec{x})\right)
\ee
have been worked out in Ref.~\cite{Baier:2007ix}. In the sound channel
we have\footnote{An analogous argument can be carried out in the shear channel
  and leads to identical conclusions.}
\bel{disprel}
\omega^3 + \f{i}{\tpi} \omega^2 - \f{k^2}{3} \left(1 + 4 \f{\eta/s}{T\tpi} \right) \,\omega - \f{i k^2}{3 \tpi} = 0
\ee
For small $k$ one finds a pair of hydrodynamic modes (whose frequency tends
to zero with $k$)
\bel{hydrom}
\omega_H^{(\pm)} = \pm \f{k}{\sqrt{3}} - \f{2 i}{3 T} \f{\eta}{s} k^2 + \ldots
\ee
and a nonhydrodynamic mode
\bel{nonhydrom}
\omega_{NH} = - i \left(\f{1}{\tpi} - \f{4}{3 T} \f{\eta}{s} k^2\right) +
\ldots
\ee
The dominant mode at long wavelengths is the one whose imaginary part is
largest (least negative).
\begin{figure}[ht]
\center
\includegraphics[height=0.4\textheight]{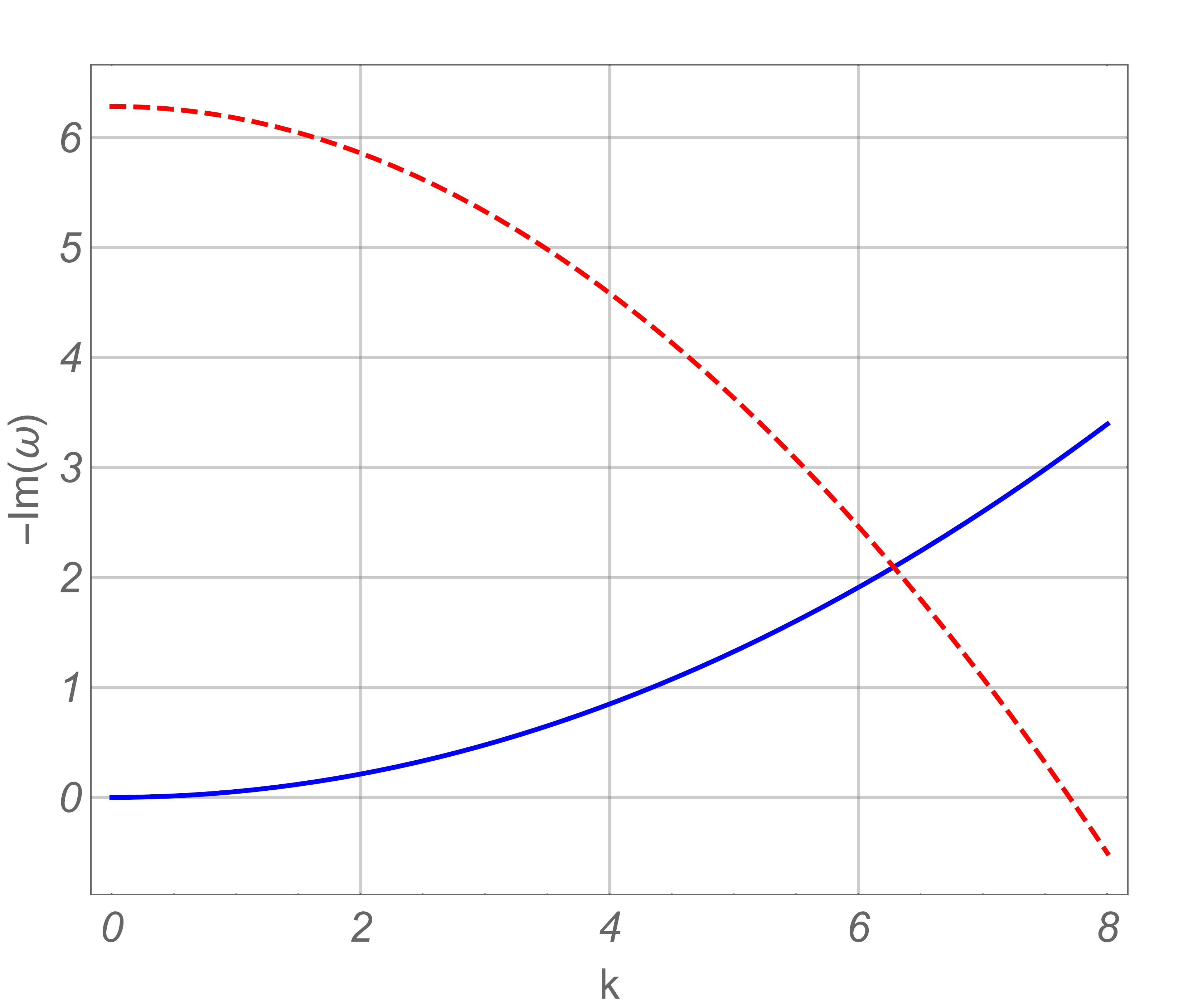}
\caption{The hydrodynamic mode (blue) and the nonhydrodynamic mode (red,
  dashed) cross at the value of $k$ given in \rf{kmax}. The plot was made
  taking $\tpi T = 2 \eta/s$ and $\eta/s=1/4\pi$. }
\label{fig:cross}
\end{figure}
One usually assumes that the hydrodynamic modes dominate, but (as seen in
\rff{fig:cross}) this is true only for $k< K$, where\footnote{This follows by
  taking the approximate solutions given in Eq.~(\ref{hydrom}) and
  (\ref{nonhydrom}). If exact solutions of \rf{disprel} were used, the
  curves in \rff{fig:cross} would not actually cross, but approach each other
  to coincide asymptotically for large $k$. The adopted procedure is an
  estimate of the scale at which the hydro modes cease to dominate.}
\bel{kmax}
K \approx \f{T}{\sqrt{2 (T\tpi) (\eta/s)}} \ .
\ee
This can be read as follows \cite{Landsteiner:2012gn}: if a drop of plasma has
spacial extent\footnote{In the context of QGP created in colliders, $R$ should
  be identified roughly with the transverse size of the plasma drop. This size
  varies very slowly in comparison with the rate of longitudinal expansion.}
$R$, then for nonhydrodynamic modes to be subdominant one would like
$R>2\pi/K$. This way one gets the condition
\bel{bound}
R T > 2\pi \sqrt{2 (T\tpi) (\eta/s)} \ .
\ee
Note that in a conformal theory $T\tpi$ as well as $\eta/s$ are dimensionless
constants.

The implication of this mode crossing\footnote{Similar mode-crossing phenomena
  have recently been discussed (in different contexts) by Janik et
  al. \cite{Janik:2015iry,Janik:2016btb}, Romatschke \cite{Romatschke:2015gic}
  and Grozdanov et al. \cite{Grozdanov:2016vgg}. In the last two references,
  the hydro modes disappear altogether rather than become subdominant. } is
that once the inequality \rf{bound} is violated, one is no longer testing
hydrodynamics, but rather the particular theory of the nonhydrodynamic sector
implicit in BRSSS theory; in other words, one is testing the regulator. In the
latter case one should really compare the regulator implicit in BRSSS theory
with alternatives, such as (for example) those discussed in
Ref.~\cite{Heller:2014wfa}.

The bound \rf{bound} is intuitively very clear and entirely consistent with
the idea that hydrodynamics may work well even in small systems as long as
they are strongly coupled, since for such systems one expects both $\eta/s$
and the relaxation time to be small. It is interesting to examine just how
small the RHS of \rf{bound} can be. The magnitude of the relaxation time is
bounded below by causality, as in \rf{causal}. There appears to be no firm
bound for $\eta/s$, but if we take the
Kovtun-Son-Starinets~\cite{Kovtun:2004de} value $\eta/s=1/4\pi$ as a
reasonable estimate, we find the simple result
\bel{minrt}
R T > 1 \ .
\ee
This inequality is reminiscent of the condition that the system size should exceed the mean free path (set by
the inverse of the temperature), but we have obtained it here without any reference to a particle description,
which may or may not exist in a given situation. The number appearing on the RHS of \rf{minrt} is the smallest
sensible minimum, which is attained using parameter values suggested by the AdS/CFT description of strongly
coupled \sym\ supersymmetric Yang-Mills theory.  In reality, the values of the relaxation time and shear
viscosity may be larger, which would imply that the nonhydrodynamic sector becomes important already on larger
scales.

The fact that the applicability of conformal hydrodynamics (in one sense or another) is governed by the
magnitude of the quantity $RT$ has been emphasized already  in Refs.~\cite{Shuryak:2013ke,Basar:2013hea}. It
is also amusing to note that our findings are consistent with Chesler's
observations~\cite{Chesler:2015bba,Chesler:2016ceu} made in the context of AdS/CFT
simulations~\cite{Chesler:2015wra}. He found that the exact, numerically calculated energy-momentum tensor can
be well approximated by hydrodynamics down to drop sizes of order $RT\approx 1$ or even somewhat less. The
analysis presented here is very different, as it refers only to the effective, hydrodynamic description, but
it is perhaps not so surprising that the same answer appears, since hydrodynamics is a very general framework,
which clearly includes systems which are strongly coupled and whose typical excitations do not have a
quasiparticle interpretation.

\section{Relation to observables}
\label{sec:pheno}


The limit \rf{minrt} can be translated into an explicit estimate of the
minimum entropy per unit of rapidity $Y$ below which one can expect regulator
independence. First note that if one neglects the transverse expansion of the
plasma drop, and follows essentially the Bjorken model \cite{Bjorken:1982qr}
(see e.g. \rfr{Florkowski:2010zz}) one has
\bel{dsdy}
\f{dS}{dY} = \pi R^2 \tau_H s \ ,
\ee
where $s$ is the entropy density and $\tau_H$ is the earliest time when
hydrodynamics could be applicable (the ``hydrodynamization time'').  Numerical
studies of thermalization based on the AdS/CFT correspondence
\cite{Heller:2011ju,Heller:2012je,Jankowski:2014lna} indicate that $w_H\equiv
\tau_H T(\tau_H)$ varies in the approximate range $0.3-0.9$ (depending on
initial conditions). These results apply directly not to QCD, but to
\sym\ supersymmetric Yang-Mills theory, but we will take them to be a
reasonable indication of the real-world situation.

To estimate the entropy density appearing in \rf{dsdy} one can take the
expression for a gas of free gas of quarks and gluons corrected by a factor of
$3/4$ to account for strong interactions. This factor can be motivated by
recalling that in \sym\ supersymmetric Yang-Mills theory the ratio of
energy density at strong coupling to the energy density at zero coupling is
$3/4$.  This way one obtains the estimate
\bel{edens}
s = \f{19}{12} \pi^2 T^3 \ ,
\ee
which is numerically very close to the result of lattice calculations
\cite{Borsanyi:2010cj} at temperatures above the deconfinement
transition.

Combining the arguments outlined above one arrives at the conclusion that
\bel{entropy}
\f{dS}{dY} = \f{19}{12} \pi^3 w_H (R_HT_H)^2 \ ,
\ee
where the subscripts indicate evaluation at $\tau=\tau_H$.
This formula can be used to translate the bound \rf{minrt} into the statement
that regulator dependence is inevitable if the entropy per unit of rapidity is
less than $(dS/dY)_{MIN} \approx 25$.

Finally, using the approximate connection between entropy and charged particle
multiplicity (see
e.g. Refs.~\cite{Florkowski:2010zz,Gubser:2008pc,Romatschke:2009im})
\bel{dndy}
\f{dS}{dY} \approx 7.5 \f{dN}{dY}
\ee
one finds
\bel{mulbound}
\left(\f{dN}{dY}\right)_{MIN} \approx 3 \ .
\ee
This result is at least qualitatively consistent with the studies of
Refs.~\cite{Romatschke:2015gxa,Habich:2015rtj}, which, as recalled in
\rfs{sec:int}, used a different, but related criterion for estimating the
limits of applicability of BRSSS hydrodynamics.

It is important to remember that to arrive
at \rf{mulbound} we assumed essentially the smallest possible values for
$\eta/s$ and the relaxation time as well as a number of other reasonable, but
not iron-clad estimates. However, the main point here is not the particular
number appearing in \rf{mulbound}, but the observation that a simple physical
argument concerning the relative importance of hydro and nonhydro modes leads
to an inequality of this kind, with the right hand side of \rf{mulbound} of
order $1$.

\section{Summary and conclusions}
\label{sec:sum}

From the perspective of fundamental theory, the reason why hydrodynamics
describes the late
time behaviour of QGP studied experimentally at RHIC and the
LHC is that it is an effective description of
the late-time behaviour of QCD. Our ignorance of QCD in this regime is
encapsulated in the values of the hydrodynamic transport coefficients which at
this time cannot be calculated {\em ab initio} and are treated as
phenomenological parameters. In principle however such a matching is
possible. Furthermore, in practice only very few of these parameters are
quantitatively relevant.

As reviewed in the Introduction, relativistic hydrodynamics necessarily
includes nonhydrodynamic modes which act as a regulator necessary for
causality. The underlying, microscopic theory, such as QCD, also has some
spectrum of nonhydro modes, but it is very difficult to make any plausible
statements about them. It is thus reasonable to focus on phenomena which do
not depend quantitatively on the details of the regulator.\footnote{In
  particular, it is hard to attempt to match the nonhydrodynamic sector of QCD
  with an effective hydrodynamic description.} The use of BRSSS theory
implicitly assumes this. As discussed in \rfs{sec:int}, experimental studies
of small systems arising in pA and pp collisions make it necessary to actually
test whether this assumption is realistic.

Here the issue was addressed directly by estimating the length scale on which
nonhydrodynamic modes decay at a rate comparable to the decay rate of hydro
modes. This leads to the inequality \rf{bound} which clearly shows that for
strongly coupled systems hydrodynamic behaviour can dominate despite small
size. This bound can be reformulated as a lower bound on the multiplicity,
supporting the conclusion of hydro simulations analysed in
Refs. \cite{Romatschke:2015gxa,Habich:2015rtj}. As discussed there, this
seemingly low bound is not at all absurd in the context of strongly coupled
systems. The basic reason for this may be summarized by saying that the
hydrodynamic description is invoked at a stage of evolution where QGP may not
even be amenable to a quasiparticle description, whereas the bound on
multiplicity pertains to the final state after the system has hadronized.

What is to be done in situations when \rf{mulbound} is violated? Since in such
cases BRSSS theory significantly depends on its implicit regulator sector, one
has to conclude that one is in fact testing this sector directly. If BRSSS
theory is found to work well in such circumstances, one should view
this as consistent with the possibility that the leading nonhydrodynamic mode
of QCD is purely damped. Otherwise one should consider using a hydrodynamic
theory with a different nonhydro sector, such as the theories proposed in
\cite{Heller:2014wfa}. In the end this may also not work unless leading
nonhydrodynamic modes are well isolated from higher ones. Whether this is true
in QCD is not known at this time.

\vskip 2em

{\bf Acknowledgments:} I would like to thank Paul Romatschke and Wojciech
Florkowski for important discussions. I also wish to thank Javier Mas and the
other organizers of NumHol2016, where this work was completed. This research
was supported by the National Science Centre Grant No. 2015/19/B/ST2/02824.


\bibliographystyle{utphys}
\bibliography{data}

\end{document}